# Identifying cybersickness causes in virtual reality games using symbolic machine learning algorithms

Thiago Porcino [a,*], Erick O. Rodrigues [b], Flavia Bernardini [a], Daniela Trevisan [a], Esteban Clua [a]

[a] Institute of Computing, Universidade Federal Fluminense (UFF), Av. Gal. Milton Tavares de Souza, s/n°, Niterói, RJ, Brazil
[b] Academic Department of Informatics, Universidade Tecnologica Federal do Parana (UTFPR), Via do Conhecimento, Km 1, Pato Branco, PR, Brazil

ABSTRACT

Virtual reality (VR) and head-mounted displays are constantly gaining popularity in various fields such as education, military, entertainment, and health. Although such technologies provide a high sense of immersion, they can also trigger symptoms of discomfort. This condition is called cybersickness (CS) and is quite popular in recent virtual reality publications. This work proposes a novel experimental analysis using symbolic machine learning to rank potential causes of CS in VR games. We estimate CS causes and rank them according to their impact using classical machine learning. Experiments are performed using two virtual reality games and 6 experimental protocols along with 37 valid samples from a total of 88 volunteers.

Our results show that rotation and acceleration triggered cybersickness more frequently in a flight game in contrast to a race game. We could also observe that subjects that are less experienced with VR are more prone to feel discomfort. Former experience plays a more important role on the race game, as this game provides more liberty to the user in terms of controllers, more displacement alternatives and a more user-controlled acceleration. Furthermore, different causes that trigger discomfort arise based on short or long term VR exposures. We suggest strategies for mitigating CS for these two scenarios: short and long term exposure experiences and compare the two highlighted scenarios (race and flight).

## 1. Introduction

Activities such as virtual training environments, simulations and entertainment in immersive virtual formats are constantly becoming more popular with the continued development and public interest in VR technologies over the last years [1]. In 2019, the VR hardware market was valued at 4.4 billion US dollars and is expected to reach 10 billion US dollars by 2022 [2].

Head-mounted displays (HMDs) is one of the means of achieving immersive virtual reality. These devices usually consist of electronic displays and lenses that are fixed over the head where the display and lenses face the eyes of the user. HMDs are used for various purposes in the industry such as in games that focus on entertainment [3], military [4], education [5], therapy [6] and simulators for numerous contexts [7].

Unfortunately, HMDs are strongly related to frequent manifestations of discomfort [8]. Among the possible manifestations, cybersickness (CS) deserves special attention as it is the most frequent and is usually associated to long exposures to HMDs. According to Ramsey et al. [9], approximately 80% of participants who have already experienced HMD-based VR reported discomfort sensations after just 10 min of exposure. In addition, more than 60% of usability problems are strongly related to discomfort [8].

The most frequent symptoms caused by CS are general discomfort, headache, stomach awareness, nausea, vomiting, sweating, fatigue, drowsiness, disorientation, and apathy [10,11]. These symptoms impact the user experience and affect the profit and coverage of the VR game industry. In addition, discomfort symptoms can vary across individuals, where some people are more susceptible than others.

Several works in the literature address the CS phenomenon and mitigation strategies for immersive VR applications using HMDs [12–15]. While most previous works [16–18] are mainly focused on detecting and predicting CS events, this work estimates and is amenable to rank the attributes that contribute the most to triggering cybersickness, enabling a selection of the most adequate strategy to mitigate CS. We propose an approach that allows the estimation of a cause while the user is under the VR condition. However, the suggestion of strategies can be implemented afterwards by the game designer. In this direction, symbolic machine learning is adequate, as understandability is essential [19].

* Corresponding author.
*E-mail address:* tmalheiros@id.uff.br (T. Porcino).
*URL:* http://www.ic.uff.br (T. Porcino).

This paper is organized as follows: Section 2 describes the literature review presenting the necessary background knowledge to understand the types of sickness associated to VR, CS strategies, how CS can be quantified, and several machine learning models to predict CS. Section 3 describes our proposed method based on symbolic machine learning and the experimental methodology with details concerning the dataset acquisition. Section 4 explains the evaluation methods and compares results for two symbolic machine learning algorithms: decision tree and random forest. Section 5 describes the identified CS causes. Section 6 presents the strategies to overcome CS. At last, Section 7 describes the conclusion, limitations, and future work.

## 2. Literature review

Motion sickness (MS) is defined as the discomfort felt during movement that is not related to body movement, e.g., in the case of airplane, boat, or land trips [20–22]. This type of discomfort also occurs in virtual environments and is called visually induced motion sickness (VIMS). The difference between VIMS and MS is the absence of real (i.e., not virtual) physical displacements that occur in MS and do not occur in VIMS [23]. Works in the literature usually label VIMS that occur in VR as CyberSickness (CS) [24]. In contrast, VIMS that occurs during flight or drive simulators is often called simulator sickness.

CS symptoms are similar to MS symptoms: nausea, vertigo, dizziness, and upset stomach [25,26]. Manifestations related to VIMS or CS may originate from various causes. Some of these causes are already properly described in the literature, whereas others still lack proper understanding and description.

However, according to the literature [27–29,8,15], it is possible to highlight the main factors that contribute to the manifestation of CS symptoms and their minimization strategies. Table 1 summarizes some strategies found in the literature .

**Table 1**
Strategies to overcome cybersickness.

| Authors | Strategies |
|---|---|
| Langbern (2018) [30] | Teleporting |
| Farmani (2018) [31] | Tunneling |
| Sapuri (2017) [32] | Trigger Walking |
| Berthoz (1975) [33] | Haptic Feedback |
| Pavard (1977) [34] | |
| Bouyer (2017) [35] | |
| Plouzeau (2018) [36] | Changes on acceleration |
| Kemeny (2017) [37] | Headlock |
| Skopp (2013) [38] | Holosphere |
| Cirio (2013) [39] | Trajectory Visualization |
| Budhiraja (2017) [40] | Rotational blur |
| Carnegie (2015) [41] | |
| Porcino (2016, 2017) [42,43] | |
| Konrad (2017) [44] | |
| Padmanaban (2017) [45] | DoF Simulation |
| Waveren (2016) [46] | Async. Time Warping for Latency |
| Kim (2012) [47] | “Cabin” Static Frame |
| Sharples (2016) [48] | |
| Kim (2017)[49] | Slow-motion |
| Bolas (2017) [50] | Dynamic FoV |
| Norouzi (2018) [51] | Dynamic Vignetting |
| Hillaire (2008) [52] | Amplified Movements |
| Plouzeau (2018) [36] | |
| Hillaire (2008) [52] | Blur Effects |
| Melo (2018) [53] | Exposure time interval |
| Lin (2004) [28] | Preparing the user visual motion |

### 2.1. Cybersickness

Unfortunately, it is not easy to numerically estimate CS as there is a lack of standardization in the literature. In addition, VR users may experience multiple symptoms and some adverse effects that are still not properly described. Another issue is the notable variation in terms of susceptibility when it comes to individuals.

The most popular way to measure CS in VR is to use subjective personal data recorded via questionnaires, such as Simulator Sickness Questionnaire (SSQ) [54] and the Motion Sickness Susceptibility Questionnaire (MSSQ) [55,56]. However, Kim H. et al. [57] revised and modified the traditional SSQ, proposing the Virtual Reality Sickness Questionnaire (VRSQ). The new VRSQ has nine items split into two classes of symptoms called “oculomotor” and “disorientation.” A recent article [58] has adhered to VRSQ. In this sense, Sevinc et al. [59] stated that SSQ is not suitable for VR applications due to its psychometrical quality issues.

VR user profile data, such as gender, age, health condition, VR experience, and psychic fatigue, are associated to the manifestation of discomfort. Women and men process visual information differently [60]. According to Biocca et al. [60], women are more inclined to MS manifestations. According to Kolasinski et al. [8], this issue happens due to a difference in the peripheral view. Women usually have a wider FoV, which increases the likelihood of discomfort. Age is another factor that can lead to an increased incidence of CS or MS. Several studies [61–63] showed that older participants were more susceptible to MS than younger ones.

Health conditions can contribute to increased susceptibility to MS or CS. According to Frank et al. [64] and Laviola et al. [65] symptoms such as stomach pain, flu, stress, hangover, headache, visual fatigue, lack of sleep, or respiratory illnesses, can lead to increased susceptibility to visual diseases.

Previous exposure to MS experiences are key in terms of discomfort susceptibility [66,67]. Individuals that are more frequently exposed to activities that trigger MS (e.g., driving, playing electronics games, etc) are less susceptible to discomfort. This is most probably due to their ability to predict future scenarios and situations in these environments [68].

### 2.2. Machine learning

Several studies use deep learning models such as convolutional neural networks (CNNs) and recurrent neural networks (RNNs) to predict CS. Kim, J. et al. [17], proposed a deep learning architecture to estimate the cognitive state using brain signals and how they are related to CS levels. Their approach is based on deep learning models, such as long short-term memory (LSTM) RNN and CNN [69–71]). The algorithms learn the individual characteristics of the participants that lead to the manifestation of CS symptoms when watching a VR video.

Jin et al. [16] grouped CS causes as follows: hardware characteristics (VR device settings and features), software characteristics (content of the VR scenes), and characteristics based on the individual user. The authors used classifiers to estimate the level of discomfort. A total of three machine learning algorithms (CNN, LSTM-RNN, and support vector regression [72]) were used. According to the authors, the LSTM-RNN obtained the best results.

Jeong et al. [18] focused on 360° VR streaming content. They analyzed scenarios of CS using brain signals. Their work uses data from 24 individuals and aims to identify common brain characteristics of VR stream patterns associated to CS manifestation. They examined VR segments and highlighted them when several individuals felt discomfort at the same time. However, the authors were not able to find specific and individual CS causes.

Garcia-Agundez et al. [73] focus on the classification of the level of CS. The proposed model uses a combination of bio-signal and game settings. User signals such as respiratory and skin conductivity of 66 participants were collected. Authors obtained a classification accuracy of 82% (SVM) when it comes to binary classifications and 56% (kNN) for the ternary case.

Moreover, Porcino et al. [74] proposed an approach that is amenable to the prediction of CS during the gameplay. Authors were able to achieve an average accuracy of 96.54% with random forest considering a total of 16 different machine learning algorithms and different scenarios. In addition, they identified attributes responsible for painful states in VR games.

## 3. Proposed method

Kim, J. et al. [17] and Jeong et al. [18] capture data using external medical equipment. This equipment is not mainstream in terms of VR. In this work, we focus on data captured without the use of extra accessories. Hence, we discard the use of any external medical equipment that could harm the user experience, or even that decrease the likelihood of the user owning the equipment in the first place.

Furthermore, Garcia-Agundez et al. [73] and Porcino et al. [74] do not focus on the estimate of the weight or on the influence of the attributes (i.e., the cause) leading to CS, as opposed to the proposal of this work.

In this novel approach, we use symbolic machine learning to analyse and identify one or more causes of discomfort, which is user and context specific. In other words, the approach described in this manuscript is not a general rule for recognizing the presence of discomfort as previously approached in the current literature. In contrast, it provides real-time user and context-sensitive evaluation and estimation of causes for cybersickness.

Moreover, the use of symbolic classifiers is paramount for an appropriate analysis and understanding of the decision. Although previous work suggest that deep learning classifiers are the most suitable approach for CS prediction [18,17], deep neural networks are black boxes that are very difficult to grasp. For this reason, this research limited the analysis to symbolic machine learning algorithms that enable a straight understanding of the decision path.

This work is built upon the understandability of *symbolic classifiers*, whose prediction process can be represented by a set of unordered or disjoint rules. Symbolic classifiers are not novel and they have been used in many scenarios where clear logical comprehensiveness is required [75]. Among the comprehensible realm of classifiers we can highlight decision tables, bayes classifiers and most notably in terms of accuracy rates, decision tree approaches. Decision tree approaches offer a robust prediction and a wide variety of algorithms and implementations in the literature [76]. A decision tree can be written as a set of disjoint unordered rules [77,78].

The logical prediction path of the decision trees inherit a personal fingerprint associated to attribute weights. Usually, attributes that are closer to the tree root are more important, as they often reduce the chaos in data more than the rest, i.e., they separate the information more appropriately than other features, improving the information gain and reducing entropy. As a general rule, the frequency in which attributes appear in the decision path is also an important piece of information. We combine these two aspects to estimate the importance of the attribute (i.e., the most important causes of discomfort).

Let us suppose a decision tree described by 13 decision nodes, as shown in Fig. 1, which contains in their conditions the features Gender, Rotation, VR experience, and Acceleration. Furthermore, let us consider a path for an instance that was predicted as discomfort, where the green arrows illustrate a single decision path.

The coverage of each node is also shown in Fig. 1. We can observe that the lower the depth of the three, the lower the coverage of each node. We used this aspect to calculate the importance of the features. In this sense, we compute a potential-cause score (PCS) by summing up the heights of these features (e.g., for a specific instance, as highlighted with the green arrows). In this case, gender appeared once and has height 3, VR experience has height 2, and Rotation has height 1. Next, the output is divided by the sum of all depths of the tree, as follows: gender = 3/6 or 50, VR experience = 2/6 or approximately 33 and Rotation = 1/6 or approximately 17. In this case, we estimate gender as the most relevant cause for CS.

Fig. 2 shows the pipeline of the proposed approach. When a model is outputted in Fig. 2(a), it is used as the "Trained Model" in Fig. 2(b) on the input dataset $S_u$, composed by the profile data and gameplay data of a user $u$. It is worth noticing that $S_u$ is composed by $N_u$ instances related to the collected $X$ features. If all instances in $S_u$ are classified as 0 (no discomfort), the flow goes to the output classification "No Discomfort". In case a subset of instances is classified as 1, the classification for this user is "Discomfort" and in this case we compute a PCS.

As a complementary remark, we use the random forest and decision tree algorithms from scikit-learn [79] python library (Python version: 3.7.5, sci-kit learn version: 0.22.1). Next section addresses the performed experiments.

### 3.1. Protocols and experiments

Two different games [74] were used for this work: (1) a race game and (2) a flight game, shown in Fig. 5. In our pipeline, we require the participants to fill in questionnaires (CSPQ [74] and two VRSQ [57]) before and after participating in a 5 or 20 min VR game session.

In the race game, the acceleration varies according to the choice of the user (they push the acceleration according to their will). In contrast, the flight game simulates an almost-constant acceleration. The player experience with both games is detailed in Fig. 3.

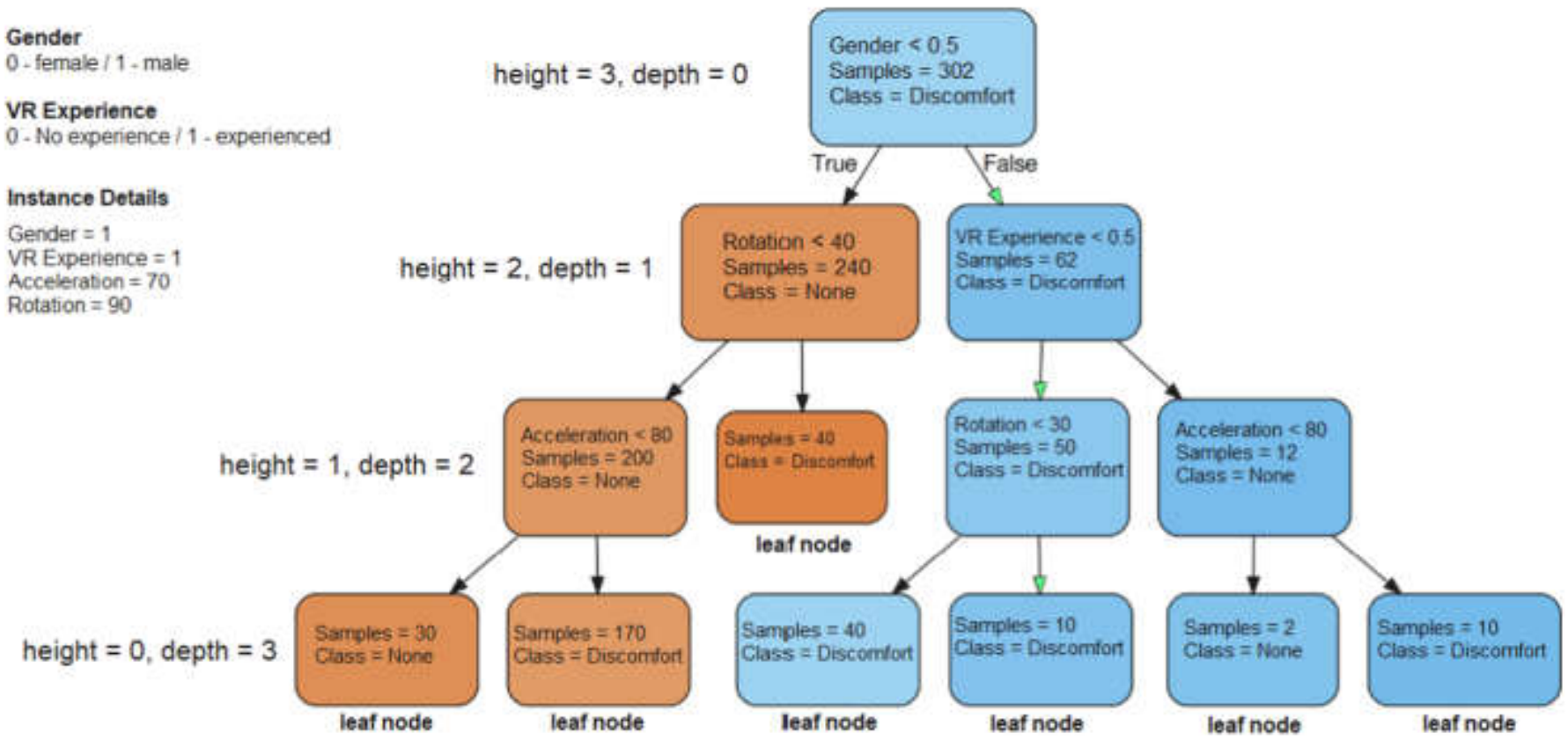

**Fig. 1.** A decision tree model. The green arrows illustrate a decision path for a specific instance that uses the attributes Gender, VR experience, and Rotation. Each attribute is associated to height values 3, 2, 1, and 0, respectively.

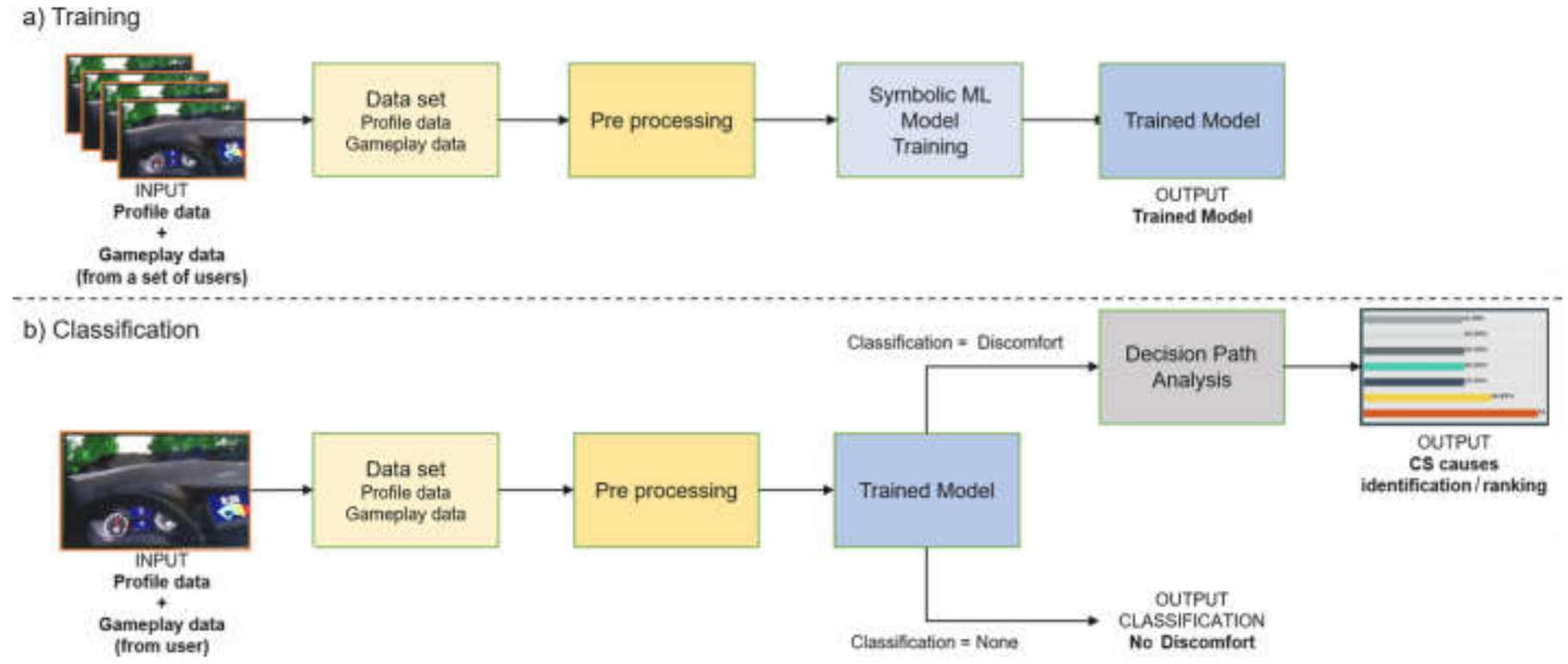

**Fig. 2.** The pipeline of the proposed approach.

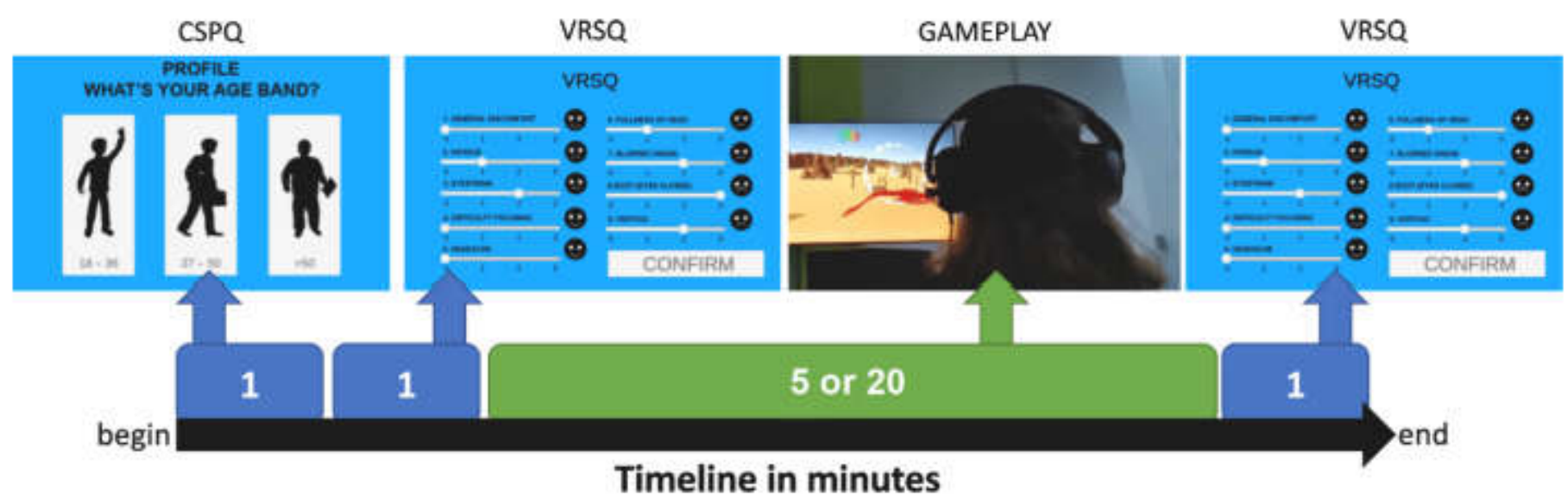

**Fig. 3.** The experience begins with the participants filling the Cybersickness Profile Questionnaire (CSPQ), followed by the VRSQ. Next, participants play the game for 5 (or 20) minutes. Finally, they fill a final VRSQ.

The data collection occurred in a few different places such as schools, universities, and technology fairs. We spent a total of tree months collecting data with two HMD devices (HTC Vive and Oculus Rift). All subjects agreed with their anonymous participation in the study and signed consent forms. The participants were allowed to quit the experiment whenever they wanted.

An protocol that was incrementally improved was used to gather all the information. We were learning along with the experience and improving the data collection protocol accordingly. Gameplay data was recorded at every second according to the user CS feedback captured by voice command. These steps and the knowledge acquired during this process are shown in the following six protocols of data collection (P1, P2, P3, P4, P5, and P6) in Table 2.

**Table 2**
Six experimental Protocols (P1, P2, P3, P4, P5, and P6) and the captured features in each of them. In P6, for safety reasons, after the complete exposure, we provided 5 min of adequate rest for each participant to self-recover from CS symptoms.

| Features | P1 | P2 | P3 | P4 | P5 | P6 |
|---|---|---|---|---|---|---|
| **Gender identification** | | | | | | |
| Male | 3 | 3 | 5 | 30 | 26 | 2 |
| Female | 1 | 1 | 1 | 7 | 9 | 0 |
| **Age range** | | | | | | |
| 18 to 36 | 2 | 4 | 6 | 33 | 28 | 0 |
| 37 to 50 | 2 | 0 | 0 | 4 | 4 | 2 |
| above 50 | 0 | 0 | 0 | 0 | 3 | 0 |
| **Used hardware** | | | | | | |
| Oculus Rift CV1 | 4 | 4 | 6 | 37 | 15 | 2 |
| HTC Vive | 0 | 0 | 0 | 0 | 20 | 0 |
| **Game** | | | | | | |
| Race | 4 | 0 | 3 | 12 | 15 | 1 |
| Flight | 0 | 4 | 3 | 25 | 20 | 1 |
| **Exposure in minutes** | | | | | | |
| Race | 5 | 5 | 5 | 5 | 5 | 20 |
| Flight | 5 | 5 | 5 | 5 | 5 | 20 |

At the end of protocol 6, a total of 37 valid users (9 women, 29 men) with ages ranging between 18 and 60 answered all the questionnaires correctly and completed the whole game interaction.

Each subject was required to complete four tasks in P5 and P6, as illustrated in Fig. 4.

In summary, we monitored the CS manifestation during the whole participant experience, which means: before, during, and after the gameplay. While we used VRSQ to monitor CS before and after the gameplay, during the gameplay, we monitored the participant verbal feedback (we registered timestamps with numbers from 0 to 3, which represents the discomfort from none to severe without interrupting the gameplay using the HMD's microphone).

The preliminary dataset contains 28 attributes obtained from the following sources: profile data and gameplay data (mentioned in [74]). All the features were captured considering two dependent variables: type of hardware and type of game. The complete list of recorded parameters are found in Table 3.

### 3.2. Data analysis

We used two combined VRSQ [57] questionnaires to compare the user discomfort level with the feedback collected from the user microphone. We calculate the VRSQ scores from each participant before and after the gameplay. The final VRSQ score is obtained with the Eq. 1, as follows:

$$VRSQ^{(final)} = |VRSQ^{(after)} - VRSQ^{(before)}| \quad (1)$$

Some subjects from P5 (8 from the race game and 12 from the flight game) were associated to zero discomfort according to the VRSQ. These cases occurred because the scores in Fig. 6 were obtained from the VRSQ output values.

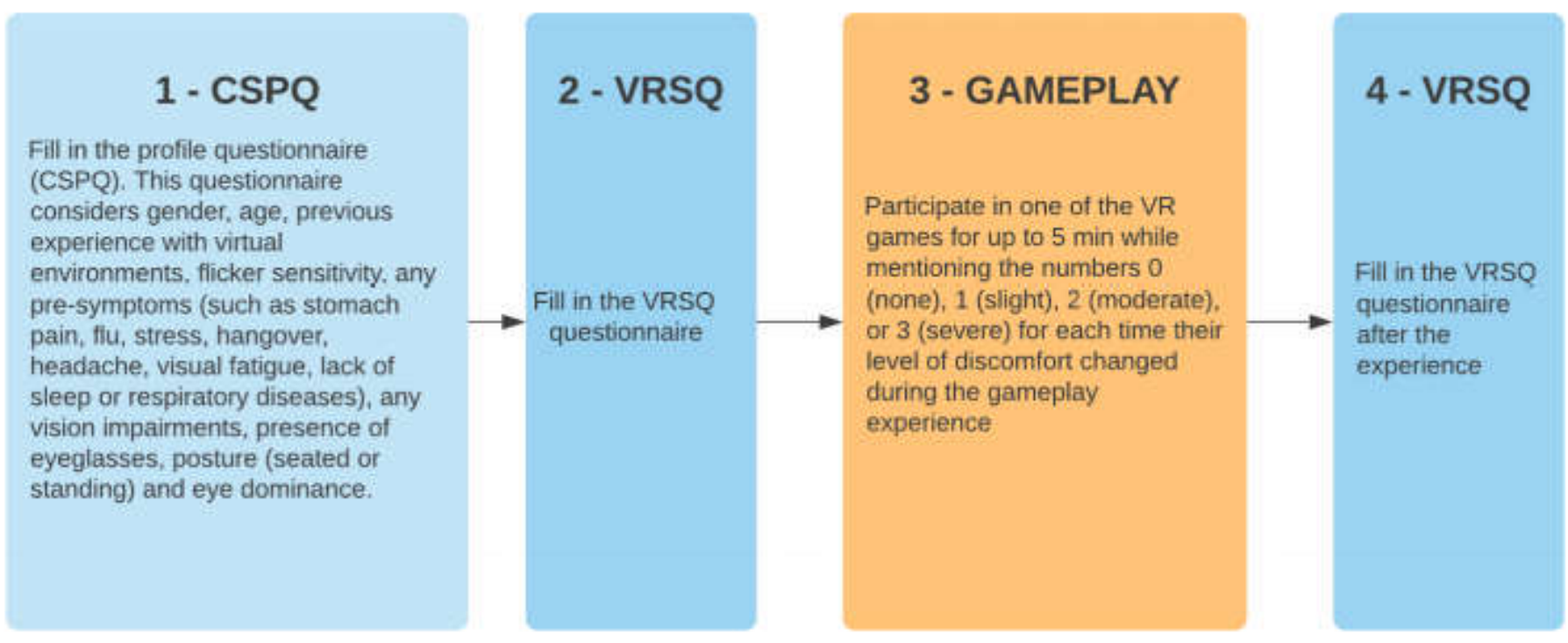

**Fig. 4.** Participant task diagram. Each participant completed four tasks during the experiment.

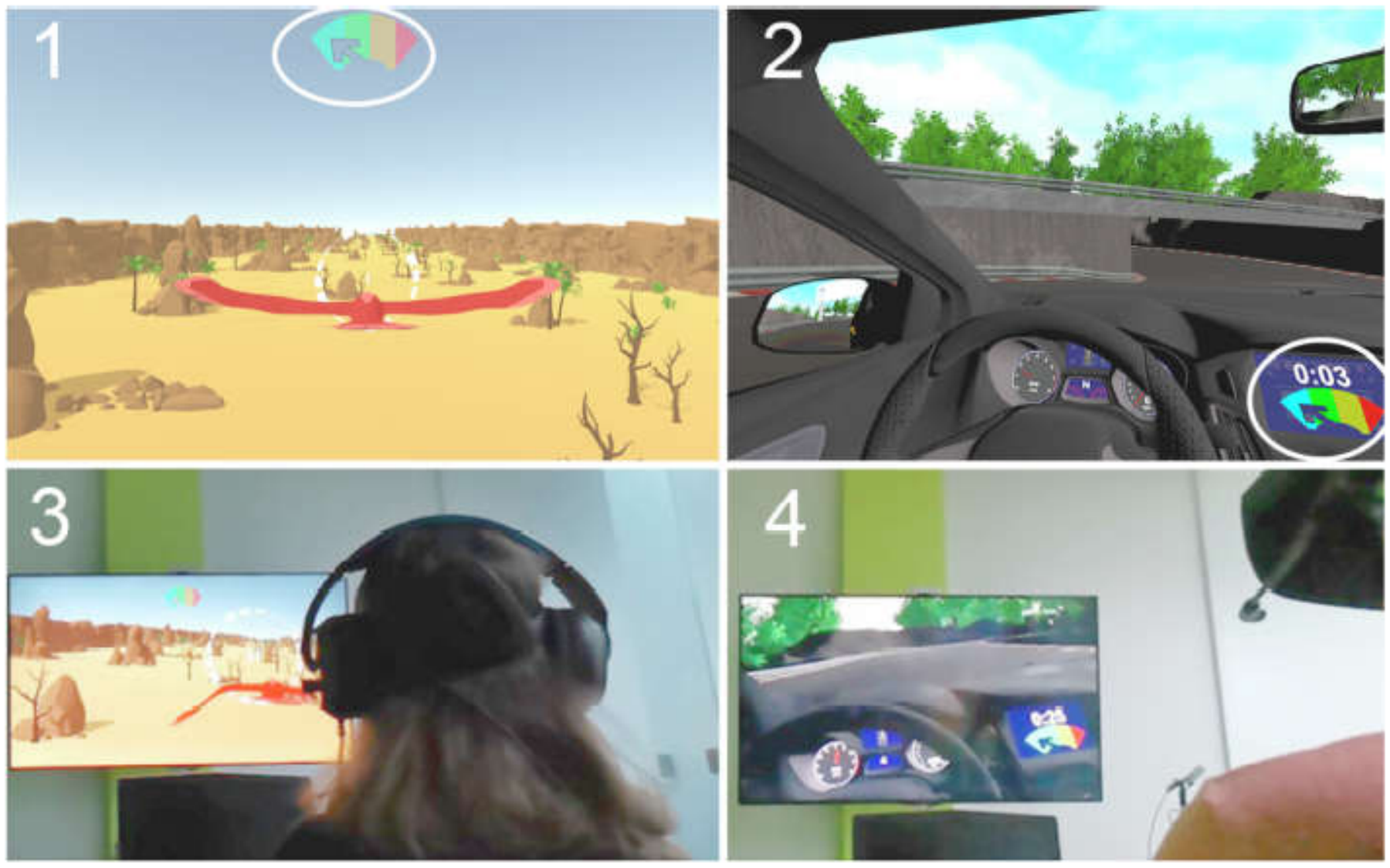

**Fig. 5.** Subjects playing the flight game at the left (1 and 3) and the race game at the right (2 and 4). In images 1 and 2, it is possible to see a visual feedback of their current discomfort level, marked with a circle.

**Table 3**
Raw feature set captured in protocol 5 and 6, which contains numerical (N) and categorial (C) features.

| Preliminary Feature set | | | |
|---|---|---|---|
| Gameplay data | | Users data | |
| Feature | Type | Feature | Type |
| Time Stamp | N | Gender | C |
| Speed | N | Age | C |
| Acceleration | N | VR Experience | C |
| Rotation (x, y and z) | N | Flicker sensitivity | C |
| Position (x, y and z) | N | Pre-symptoms | C |
| Region Of Interest | C | Glasses wearing | C |
| Size of FOV | N | Vision Impairments | C |
| Frame Rate | N | Posture | C |
| Static Frame | C | Dominant Eye | C |
| Haptic Feedback | C | Discomfort Level | N |
| Degree of Control | C | | |
| DoF Simulation | C | | |
| Player Locomotion | C | | |
| Automatic Camera | C | | |

- The subject immediately felt better when removing the HMD equipment, and for this reason, the subject replied as “feeling no discomfort”.
- The subject did not experience discomfort during the experiment.
- The subject marked the questionnaire with no care.

To validate each participant data, we compare the results from VRSQs to the results obtained with verbal feedback. In other words, we interpret the results following this reasoning: if a participant scored positive in VRSQ (meaning discomfort) but does not highlight the discomfort verbally (meaning no discomfort), we considered the user as unreliable data and discarded them from the machine learning analysis. For this reason, the further experimental analysis of this work (ranking of causes) considers just the data from users whose VRSQ scores were positive. This translates to a total of 17 participants out of the 37 mentioned before (Fig. 6 and Table 4). Specifically, at each second of gameplay, we collect a new instance, and in total, we recorded 11722 instances with the same attribute set. In other words, the more precise way to look at this problem is to be second-oriented, as the instance (or each second) is the object of classification in terms of prediction of CS manifestation and further analysis.

### *3.3. Preliminary analysis*

To perform the preliminary analysis of the recorded data, we created a data visualization application using Unity 3D. This application read our recorded dataset using the CSV format and create graphical visualizations using position and discomfort based on the verbal feedback.

In this analysis, we used 3993 samples from the P5 race game using position and discomfort level data presented in Table 3. We observed that discomfort occurs throughout the track (Fig. 7). However, the discomfort levels vary in specific regions.

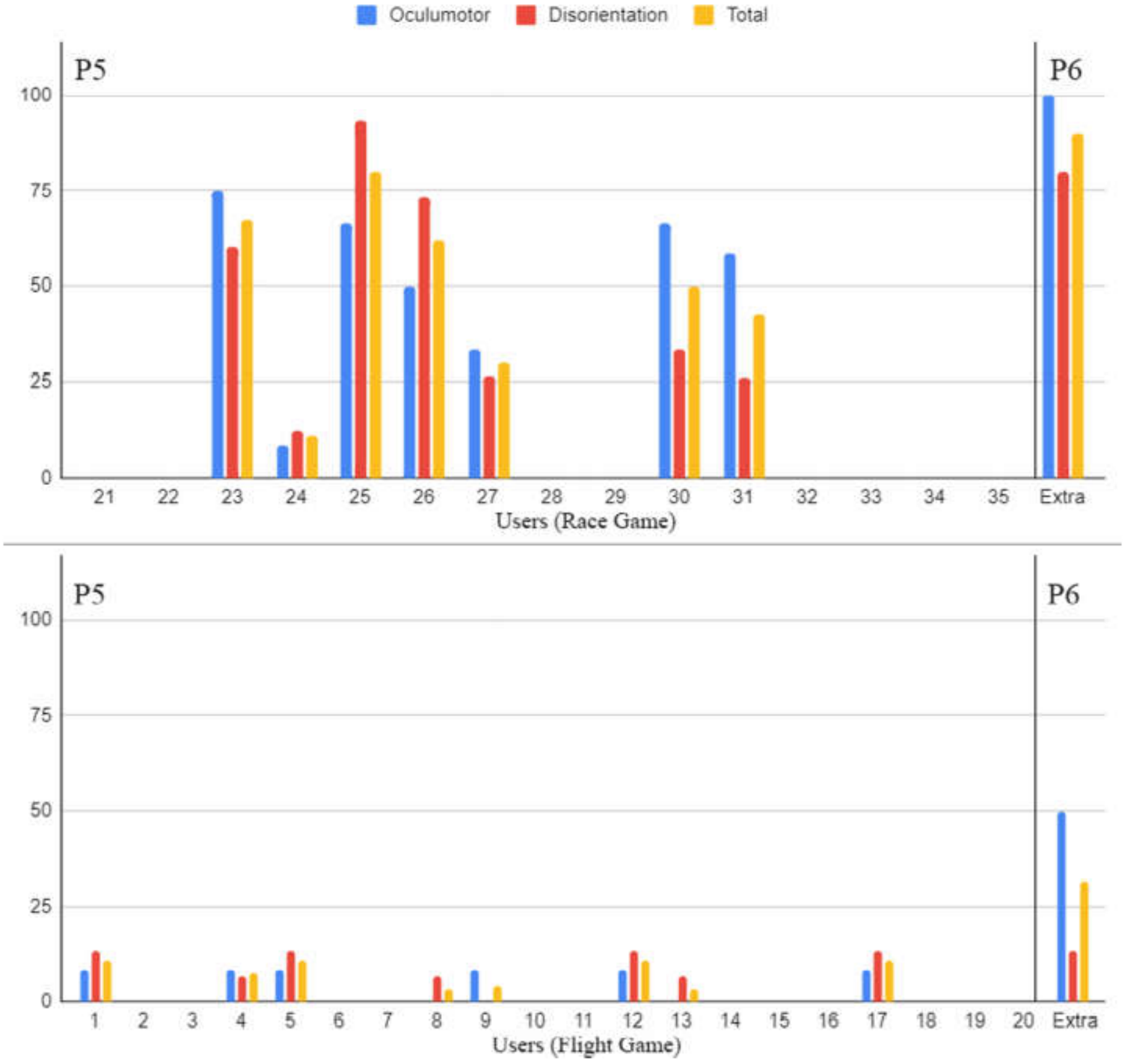

**Fig. 6.** Obtained VRSQ scores in P5 and P6. Concerning both game scenarios, 15 users achieved positive scores from P5 and 2 from P6.

In a comparative sample of reports of discomfort between individuals of the female gender (7 total) and male (8 total), we observed that in an accumulated result, the male participants reported discomfort values greater than zero more often than individuals of the female gender (illustrated in Fig. 8).

As the race game has a more stable balance between genders (7 women and 8 men), we conducted gender analysis just with the race game. Moreover, we observed that women played the race game using a more constant acceleration and moved slowly, while men played the race game in a more competitive way, varying acceleration and speed more frequently. However, this conclusion might not be statistically significant due to the number of valid participants.

### 3.4. *Pre-selection of features*

We ranked the features in terms of importance using the Weka [80] classifier attribute evaluator, choosing the "leave one attribute out" option. This algorithm removes attributes from the dataset and evaluates how its removal influences on the performance of the classification, using this importance to create the final ranking. This result is shown from top to bottom in order of importance in Table 5. (see Table 6).

For our experimental approach, we merged the top six features for each one of the scenarios (race and flight games) shown in Table 5. In addition, we discarded the position attribute. The position attribute is specific to each game and can produce overfitting or even end up producing low accuracy rates when used with games other than the ones used during the training protocol. Later, extracted features were used as training data to construct the decision tree and random forest models.

The four original classes, namely, none, slight, moderate, and severe, were converted into binary classes: 0 for no discomfort and 1 for discomfort. Class 1 translates to slight, moderate, or severe classes. In the next section, we provide the performed experiments and obtained results.

## 4. Evaluation method and results

Leave-one-out (leaving one subject out) was chosen as evaluation method, which is a particular case of the cross-validation technique where the number of folds matches the number of instances in the dataset. In other words, one subject was left out while the algorithm was trained with n-1 subjects and the model was tested with the left out subject. This process is repeated n times switching the left out subject, where n is the total amount of subjects.

We also perform an efficiency analysis in terms of the depth of the generated decision trees. For the race game the optimal tree depth was 7 and for the flight game it was 9. No significant improvements on the AUC (area under the ROC curve) scores were observed with large depths. It is important to highlight that large trees also consume greater processing time, both in the case of training and classification, and also greater memory consumption.

When it comes to the comparative results concerning the AUC scores (Tables 7 and 8), the random forest classifier obtained the best results. However, random forest is an ensemble algorithm, and hence it is naturally more complex and slower in terms of run time when compared to a single decision tree. Despite of this fact and due to its great generalization capabilities, we adopted the random forest as the classifier for the further analyses of this manuscript.

We focus on minimizing the minimum tree coverage, maximizing AUC scores and minimizing the tree depth for better efficiency. This reasoning was used to choose the ideal tree depth. The ideal depths were 7 and 9 for the race and flight game, respectively. Figs. 9 illustrate the minimum coverage as the depth of the tree is increased, respectively. Moreover, Fig. 10 illustrates the discomfort decision path construction

**Table 4**
A summary of valid recorded data in race and flight games. Concerning both scenarios, 11722 instances were captured, and 17 participants achieved positive scores for VRSQ (highlighted).

| Race Game | | | | Flight Game | | | |
|---|---|---|---|---|---|---|---|
| User | Instances | Exposure (minutes) | VRSQ (total) | User | Instances | Exposure (minutes) | VRSQ (total) |
| 21♂ | 300 | 5 | 0 | 1♂ | 103 | 1.71 | 10.83 |
| 22♂ | 104 | 1.73 | 0 | 2♂ | 300 | 5 | 0 |
| 23♂ | 300 | 5 | 67.5 | 3♀ | 300 | 5 | 0 |
| 24♂ | 300 | 5 | 10.83 | 4♂ | 300 | 5 | 7.5 |
| 25♂ | 300 | 5 | 80 | 5♂ | 102 | 1.7 | 10.83 |
| 26♂ | 300 | 5 | 61.6 | 6♂ | 300 | 5 | 0 |
| 27♂ | 300 | 5 | 30 | 7♂ | 300 | 5 | 0 |
| 28♀ | 24 | 0.4 | 0 | 8♂ | 300 | 5 | 3.3 |
| 29♀ | 300 | 5 | 0 | 9♀ | 300 | 5 | 4.16 |
| 30♀ | 287 | 4.78 | 49.9 | 10♂ | 300 | 5 | 0 |
| 31♀ | 250 | 4.16 | 42.5 | 11♂ | 300 | 5 | 0 |
| 32♀ | 300 | 5 | 0 | 12♂ | 300 | 5 | 10.83 |
| 33♂ | 300 | 5 | 0 | 13♂ | 52 | 0.86 | 3.33 |
| 34♀ | 300 | 5 | 0 | 14♂ | 300 | 5 | 0 |
| 35♀ | 300 | 5 | 0 | 15♂ | 300 | 5 | 0 |
| ExtraA♂ | 1200 | 20 | 90 | 16♂ | 300 | 5 | 0 |
| | | | | 17♂ | 300 | 5 | 10.83 |
| | | | | 18♂ | 300 | 5 | 0 |
| | | | | 19♂ | 300 | 5 | 0 |
| | | | | 20♂ | 300 | 5 | 0 |
| | | | | ExtraB♂ | 1200 | 20 | 31.6 |
| **Total** | **5165** | **86.08** | **432.33** | | **6557** | **109.28** | **93.21** |

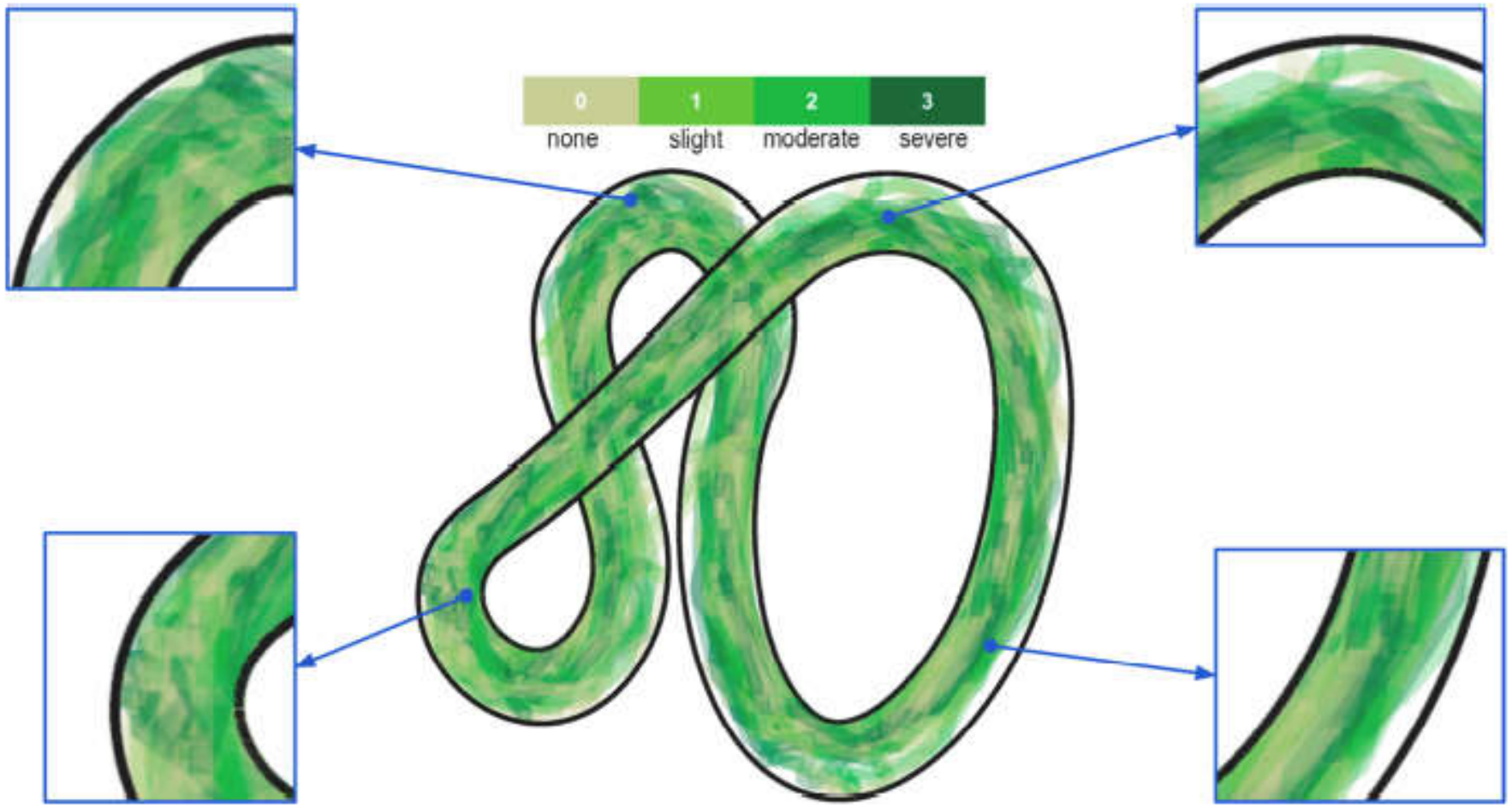

**Fig. 7.** Visualization of all occasions where the participants reported some level of discomfort during the experiment. In the image, the intensity of discomfort reported by users varies from 0 (none) to 3 (severe).

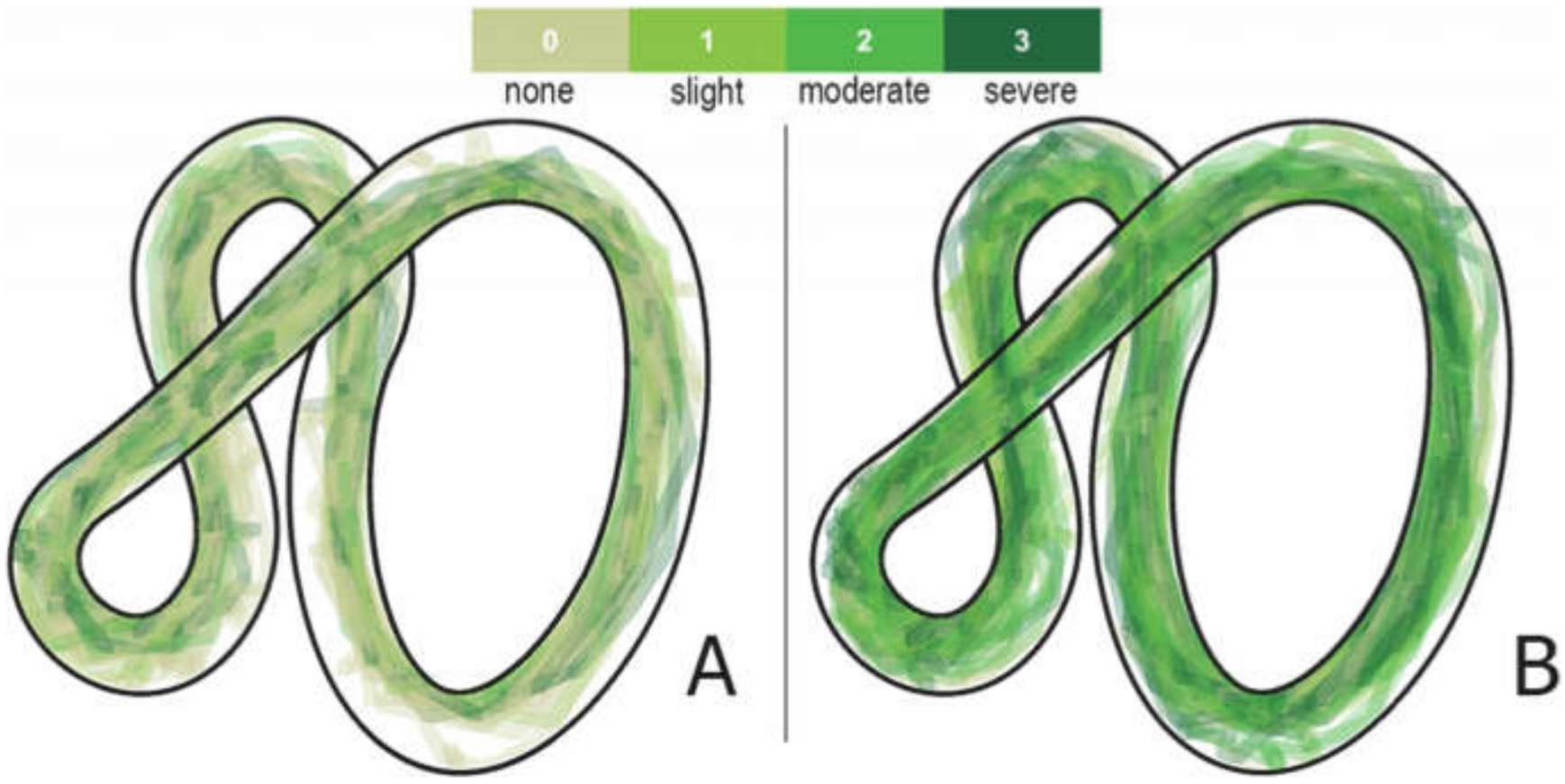

**Fig. 8.** Comparison of the discomfort reported by female and male participants. The intensities of discomfort reported by users vary from 0 (none) to 3 (severe).

**Table 5**
Weka binary classification attribute evaluation.

| Binary Classification Attribute Ranking | | |
|---|---|---|
| Rank | Race Game | Flight Game |
| 1 (best) | Time Stamp | Age |
| 2 | Age | Position Z |
| 3 | Gender | VR Experience |
| 4 | Rotation Z | Vision Imp. |
| 5 | Speed | Rotation Z |
| 6 | VR Experience | Time Stamp |
| 7 | Rotation X | Rotation X |
| 8 | Rotation Y | Speed |
| 9 | Eye Dominance | Eye Dominance |
| 10 | Region of Interest | Position Y |

**Table 6**
Final feature set.

| Gameplay data | Profile data |
|---|---|
| Timestamp | Gender |
| Speed | Age |
| Acceleration | VR Experience |
| Rotation Z | |
| Frame Rate | Discomfort Level |

**Table 7**
User-specific AUC scores in the race game (tree depth 7).

| User | Decision Tree | | Random Forest | |
|---|---|---|---|---|
| | Train | Test | Train | Test |
| 21 | 0.94 | 0.28 | 0.95 | 0.76 |
| 22 | 0.94 | 0.72 | 0.95 | 1.00 |
| 23 | 0.94 | 0.96 | 0.95 | 1.00 |
| 24 | 0.95 | 0.58 | 0.95 | 0.73 |
| 25 | 0.96 | 0.39 | 0.97 | 0.51 |
| 26 | 0.93 | 0.37 | 0.95 | 0.97 |
| 27 | 0.92 | 0.85 | 0.93 | 0.95 |
| 28 | 0.94 | 0.50 | 0.95 | 0.24 |
| 29 | 0.94 | 0.93 | 0.95 | 0.97 |
| 30 | 0.96 | 0.72 | 0.96 | 0.77 |
| 31 | 0.95 | 0.65 | 0.96 | 0.78 |
| 32 | 0.92 | 0.05 | 0.94 | 0.19 |
| 33 | 0.93 | 0.74 | 0.94 | 0.96 |
| 34 | 0.95 | 0.58 | 0.95 | 0.61 |
| 35 | 0.94 | 0.54 | 0.95 | 0.75 |
| **Average** | **0.94** | **0.58** | **0.95** | **0.77** |

**Table 8**
User-specific AUC scores in the flight game (tree depth 9)

| User | Decision Tree | | Random Forest | |
|---|---|---|---|---|
| | Train | Test | Train | Test |
| 1 | 0.98 | 0.62 | 0.97 | 0.54 |
| 2 | 0.98 | 0.99 | 0.96 | 1.00 |
| 3 | 0.98 | 0.50 | 0.96 | 0.80 |
| 4 | 0.99 | 0.77 | 0.97 | 0.83 |
| 5 | 0.99 | 1.00 | 0.97 | 1.00 |
| 6 | 0.98 | 0.95 | 0.97 | 0.99 |
| 7 | 0.99 | 0.93 | 0.97 | 0.99 |
| 8 | 0.99 | 0.81 | 0.97 | 0.95 |
| 9 | 0.98 | 0.33 | 0.97 | 0.09 |
| 10 | 0.98 | 0.82 | 0.97 | 0.95 |
| 11 | 0.99 | 0.83 | 0.98 | 0.94 |
| 12 | 0.99 | 0.49 | 0.98 | 0.26 |
| 13 | 0.98 | 1.00 | 0.97 | 1.00 |
| 14 | 0.99 | 0.77 | 0.97 | 0.93 |
| 15 | 0.99 | 0.92 | 0.97 | 1.00 |
| 16 | 0.99 | 0.75 | 0.97 | 0.82 |
| 17 | 0.98 | 0.53 | 0.97 | 0.90 |
| 18 | 0.98 | 0.48 | 0.97 | 0.46 |
| 19 | 0.99 | 0.92 | 0.98 | 0.97 |
| 20 | 0.98 | 0.55 | 0.97 | 0.98 |
| **Average** | **0.99** | **0.79** | **0.97** | **0.95** |

using depth 7 and a simple decision tree to identify CS and estimate the CS cause.

## 5. Identification of causes

Predicting the cause of CS is not trivial. Every user has a specific susceptibility to discomfort. Furthermore, several attributes are related to the hardware and ergonomic aspects of the devices. We are still far from tracing very precise causes for all specific cases. However, so far, factors such as rotation, speed, gender, and previous VR experience, appeared as dominant factors that can trigger CS.

Our approach works with up to eight factors attributed to CS. Previous works in the literature already proposed strategies for four of these attributes (exposure time, acceleration, speed, frame rate, and camera rotation on the z-axis) [53,35,40,46]. The remaining causes (gender, VR experience, and age) are causes associated to the user profile and are still not associated to a clear strategy. In addition, we observed different patterns of causes for users in the race game when compared to the flight game.

Exposure time (timestamp) was the most frequent cause of discomfort. Overall, the race game contributed more to CS manifestation (39.4) when compared to the flight game (35.9). A possible suggestion is to reduce the time of exposure in the case of race games. CS triggered by acceleration shifts controlled by users was less frequent in the race game (5.6) when compared to the flight game, where acceleration was not controlled by the user (11.80).

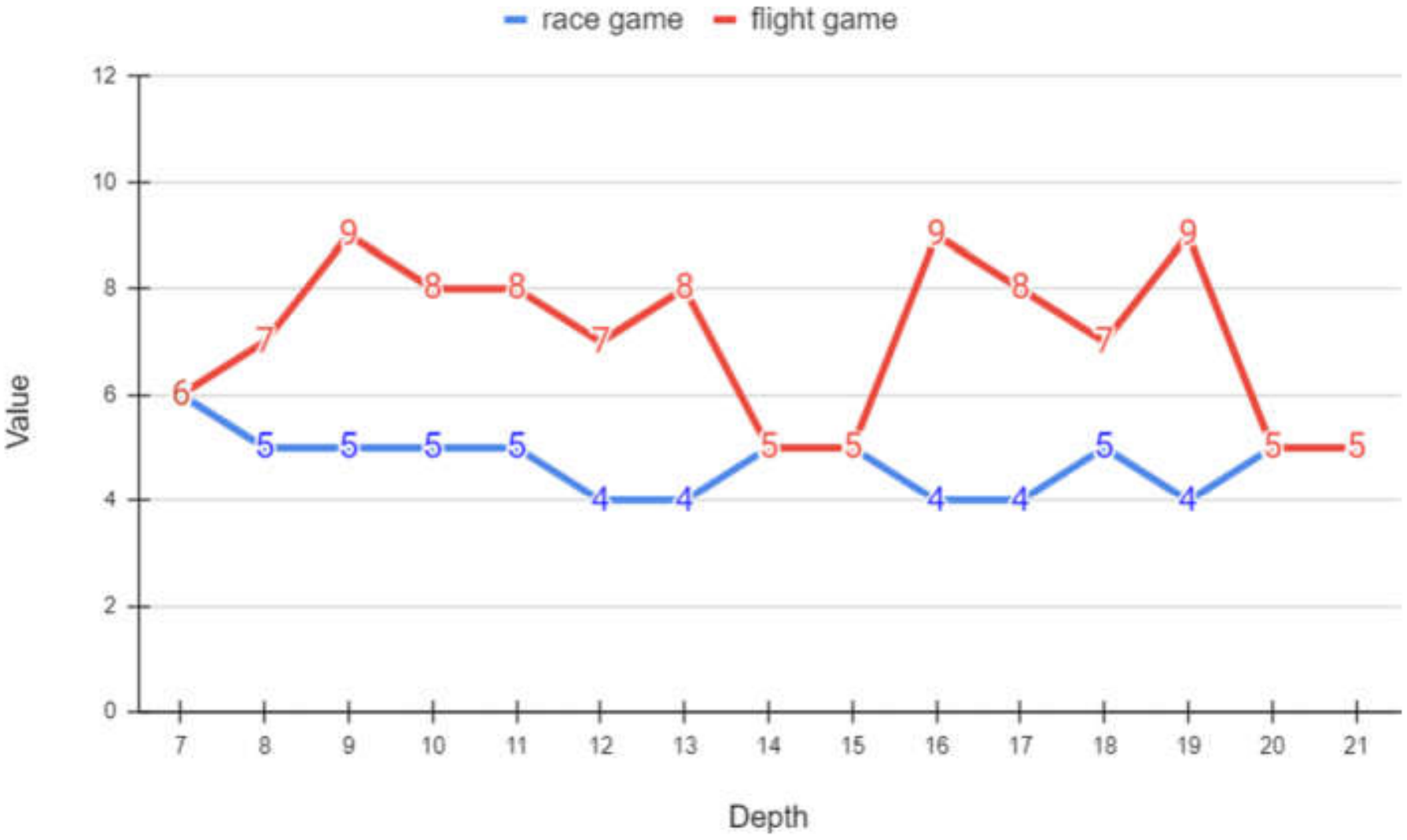

**Fig. 9.** Minimum coverage in leaf nodes for Scenarios A (race game) and B (flight game).

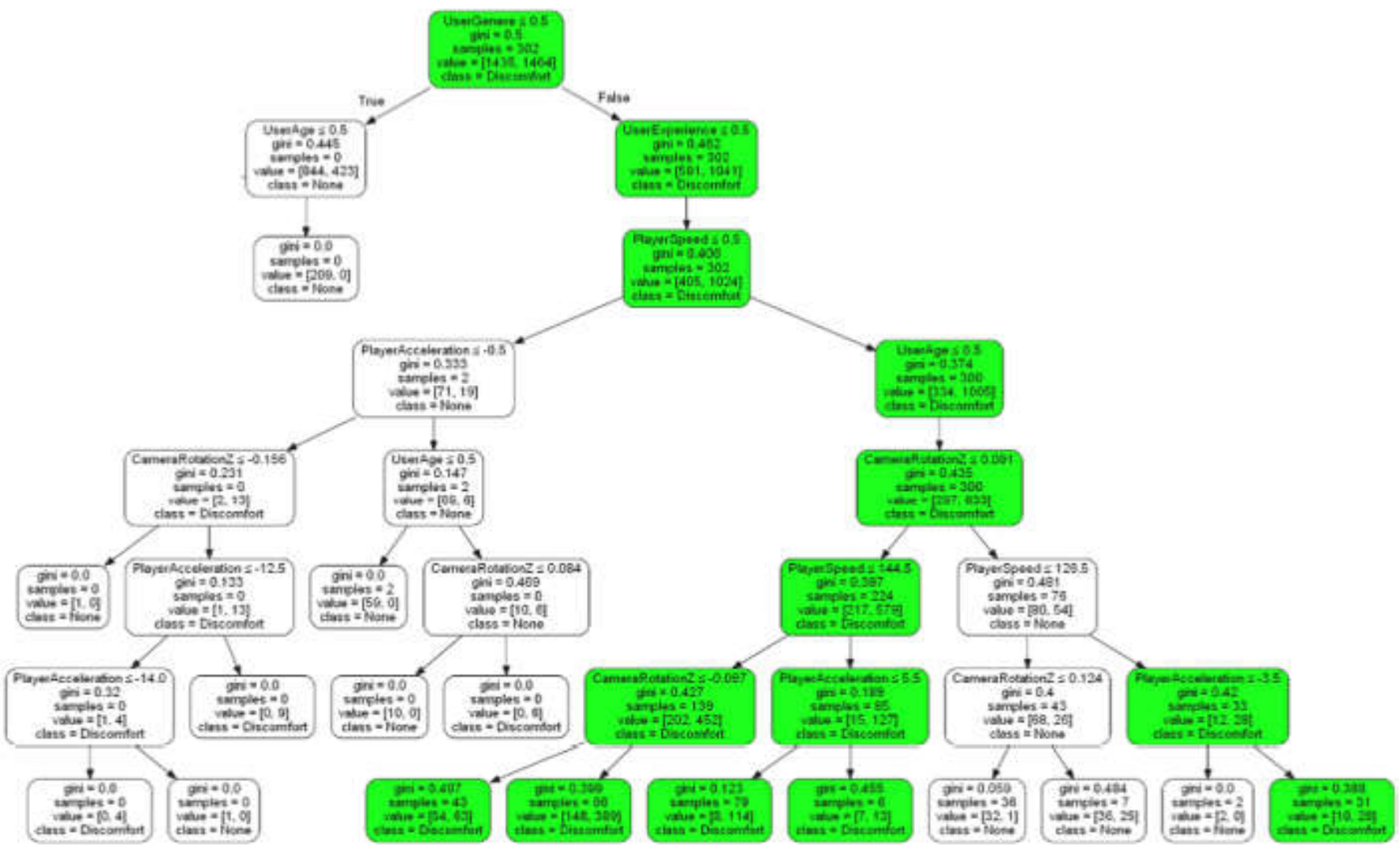

**Fig. 10.** Example of the decision tree discomfort path in the race game (depth 7). The nodes where the class is equal to discomfort (or 1) and the number of samples that are greater than zero are highlighted in green.

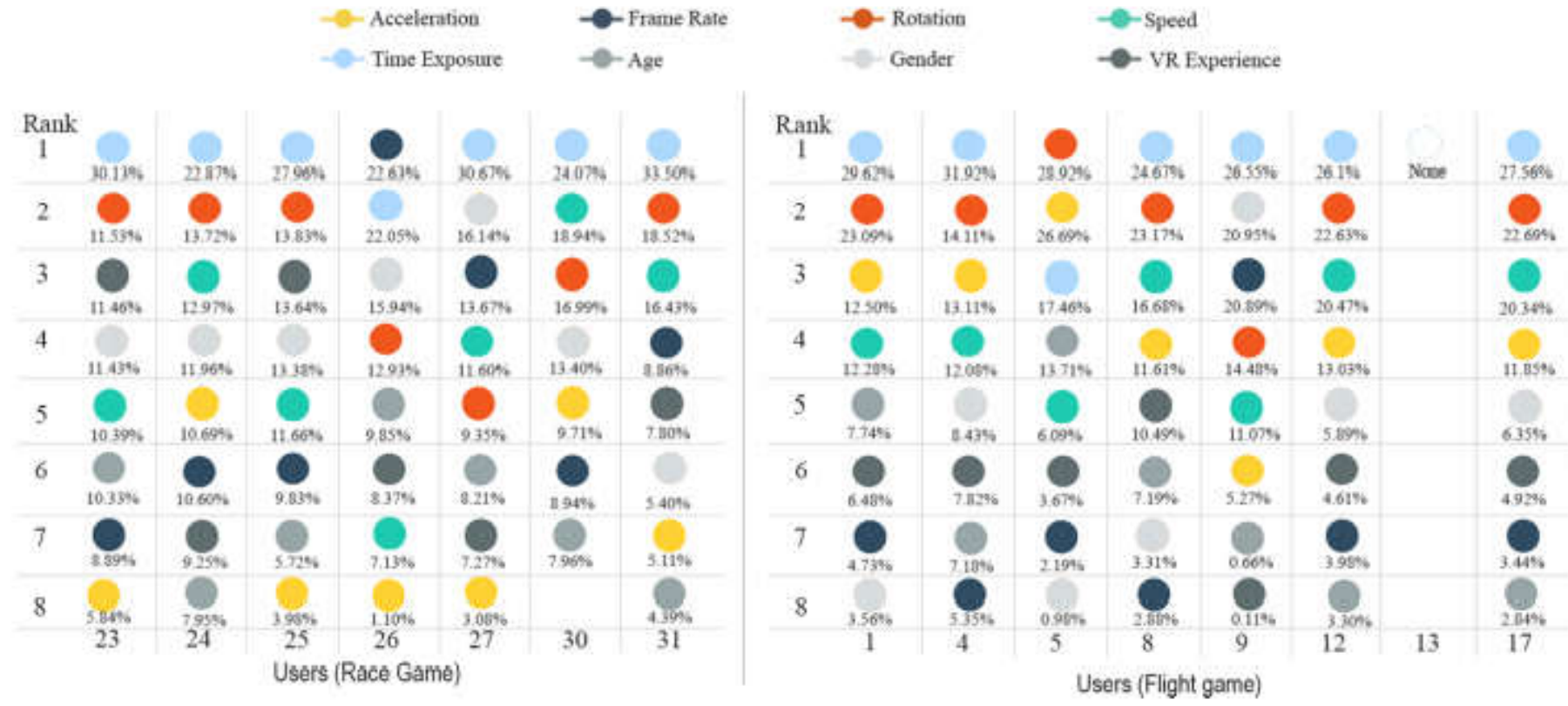

**Fig. 11.** Random Forest feature ranking (identification of cybersickness causes) for the race (A) and flight game (B) for P5 subjects.

Another feature that influences on the discomfort in both scenarios is the former VR experience. PCS was greater in the race game in contrast to the flight game, 8.25 and 4.76, respectively. In addition, rotation was marked as cause more frequently in the flight (18.70) when compared to the race game (13.83).

## 6. Discussion

We observed that most of PCS, when it comes to short-exposure cases, were tied to exposure time, rotation, and acceleration (Fig. 11). This rank of causes enabled us to associate CS minimization strategies that conform to the literature. Moreover, these strategies can be used to minimize CS problems during short or long exposures. In what follows, we analyze different periods of a 20 min long gameplay experience.

We used our approach based on random forest to predict CS. The resulting suggestion of strategies, as well as each rank of CS causes can be observed in Table 9 and Table 10. These strategies were chosen based on their obtained PCS and according to the usual strategies known to the literature.

### 6.1. Suggestion of strategies for race game

When it comes to the race game, the top rank cause of CS (Fig. 12) was exposure time, where the PCS average was equal to 32.65. This rate lead us to suggest strategies to reduce the exposure such as introducing intervals within the VR experience (more details in Table 9). (Fig. 13).

When we analyse the 20 min of gameplay of subjects from P6, the first 5 min indicate that velocity plays a very important role in the triggering of CS, achieving up to 28.97 of PCS, while later decreasing the PCS rate and ultimately achieving 8.59 at the end of the exposure. Strategies such as slow-motion [49] can be applied to minimize discomfort when velocity is the most important variable leading to the symptom. This analysis illustrates that it is possible to use our methodology to suggest strategies during long VR experiences.

We separated the 20 min long gameplay into four sections to analyse how the rotation variable behaves in terms of triggering CS. In this case, the PCS were fairly similar, 10.65, 14.26, 14.47, and 12.04, respectively, for the entire experience (Table 9).

Overall, rotation was ranked 3rd in most gameplay experiences. This finding suggests the usage of strategies to cover rotational problems in a general basis. Rotational blurring [40] and tunneling [31] can be applied specially when high velocity curves are at stake.

### 6.2. Suggestion of strategies for the flight game

As with the race game, the flight game also displays a high PCS for the exposure time variable (31.43), which is prevalent for the entire 20 min of game experience (12). The same strategy, the introduction of intervals, should also be considered in this case (more details in Table 9).

Rotation was ranked first from minute 5 to minute 10 (PCS: 25.45). For this reason, rotational blur [40] combined to tunneling [31] are appropriate choices for this game.

Acceleration achieved high PCSs from minute 5 to minute 10 (PCS: 17.47), and also from minute 15 to minute 20 (PCS: 15.52). Real time adjustments of the acceleration are a good fit for this case [36].

Overall, causes were ranked differently over the gameplay exposure. CS can be triggered by different factors and combinations of them, where eventually a single variable influences more than the rest.

## 7. Conclusions

In this work, we propose an approach to identify causes of cybersickness in different virtual reality games using head-mounted displays. To the best of our knowledge, this is the first work that uses symbolic classifiers to analyze causes of cybersickness during the gameplay experience. Once the cause is identified, game designers are able to select the most adequate strategy to mitigate the impacts of CS, according to the literature.

We experimented with two different scenarios and proposed the use of two symbolic machine learning algorithms, along with an analysis to identify the optimal tree depth for the generated models. Next, we performed a feature ranking to identify the most relevant causes of CS, which vary along with the gameplay experience and is related to the user. We observed that exposure time, rotation, and acceleration are most likely the top factors contributing to CS. As exposure time appeared as a preliminary cause for CS in our experiments, we suggest reducing the time of exposure in the case of race games (to 5 min max) or to provide intervals for the user at every 5 min of experience.

At last, we conclude that introducing rapid movements and related variables that are controlled by the user can potentially lead to higher incidence of cybersickness. Virtual reality games that rely on low complexity controllers are a better fit for non-experienced users.

As a final remark, the raw dataset of this work along with developed games are available in the following public domain: for further reproduction and comparison [81].

**Table 9**
Strategies for the race game (P6 subjects).

| | 0 to 5 min | | | | 10 to 15 min | | |
|---|---|---|---|---|---|---|---|
| Rank | Causes | PCS | Strategies | Rank | Causes | PCS | Strategies |
| 1 | Exposure time | 35.78 | Interval | 1 | Exposure time | 25.32 | Interval |
| 2 | Speed | 28.97 | Slow-motion | 2 | Gender | 20.92 | None |
| 3 | Rotation | 10.65 | Rotational blurring, Tunneling, Slow-motion | 3 | Rotation | 14.47 | Rotational blurring, Tunneling, Slow-motion |
| 4 | Gender | 9.69 | None | 4 | Frame rate | 11.91 | None |
| 5 | Age | 8.5 | None | 5 | VR Experience | 10.05 | None |
| 6 | Acceleration | 2.95 | Acceleration adjustment, Haptic feedback | 6 | Speed | 7.93 | Slow-motion |
| 7 | Frame rate | 2.78 | None | 7 | Age | 5.73 | None |
| 8 | VR Experience | 0.68 | None | 8 | Acceleration | 3.66 | Acceleration adjustment, Haptic feedback |
| | 5 to 10 min | | | | 15 to 20 min | | |
| Rank | Causes | PCS | Strategies | Rank | Causes | PCS | Strategies |
| 1 | Exposure time | 30.76 | Interval | 1 | Exposure time | 38.74 | Interval |
| 2 | Speed | 19.19 | Slow-motion | 2 | Gender | 16.23 | None |
| 3 | Rotation | 14.26 | Rotational blurring, Tunneling, Slow-motion | 3 | Frame rate | 13.7 | None |
| 4 | Age | 14.23 | None | 4 | Rotation | 12.04 | Rotational blurring, Tunneling, Slow-motion |
| 5 | Acceleration | 12.39 | Acceleration adjustment, Haptic feedback | 5 | Speed | 8.59 | Slow-motion |
| 6 | Gender | 4.37 | None | 6 | VR Experience | 6.39 | None |
| 7 | Frame rate | 2.83 | None | 7 | Age | 4.3 | None |
| 8 | VR Experience | 1.97 | None | 8 | Acceleration | 0 | Acceleration adjustment, Haptic feedback |

**Table 10**
Strategies for the flight game (P6 subjects).

| 0 to 5 min | | | | 10 to 15 min | | | |
|---|---|---|---|---|---|---|---|
| Rank | Causes | PCS | Strategies | Rank | Causes | PCS | Strategies |
| 1 | Exposure time | 33.13 | Interval | 1 | Exposure time | 28.80 | Interval |
| 2 | Speed | 17.37 | Slow-motion | 2 | Speed | 18.90 | Slow-motion |
| 3 | Acceleration | 12.48 | Acceleration adjustment, Haptic feedback | 3 | Rotation | 16.04 | Rotational blurring, Tunneling, Slow-motion |
| 4 | Rotation | 12.19 | Rotational blurring, Tunneling, Slow-motion | 4 | Acceleration | 9.70 | Acceleration adjustment, Haptic feedback |
| 5 | VR Experience | 9.21 | None | 5 | Age | 9.51 | None |
| 6 | Gender | 5.96 | None | 6 | Frame rate | 7.15 | None |
| 7 | Age | 5.85 | None | 7 | VR Experience | 6.51 | None |
| 8 | Frame rate | 3.82 | None | 8 | Gender | 3.39 | None |
| **5 to 10 min** | | | | **15 to 20 min** | | | |
| Rank | Causes | PCS | Strategies | Rank | Causes | PCS | Strategies |
| 1 | Rotation | 25.45 | Rotational blurring, Tunneling, Slow-motion | 1 | Exposure time | 40.67 | Interval |
| 2 | Exposure time | 23.13 | Interval | 2 | Acceleration | 15.52 | Acceleration adjustment, Haptic feedback |
| 3 | Acceleration | 17.47 | None | 3 | Age | 12.54 | None |
| 4 | VR Experience | 11.15 | None | 4 | Gender | 12.54 | None |
| 5 | Frame rate | 8.97 | None | 5 | Speed | 12.38 | Slow-motion |
| 6 | Age | 6.55 | None | 6 | Rotation | 9.71 | Rotational blurring, Tunneling, Slow-motion |
| 7 | Speed | 4.71 | Slow-motion | 7 | Frame rate | 7.91 | None |
| 8 | Gender | 2.57 | Acceleration adjustment, Haptic feedback | 8 | VR Experience | 1.27 | None |

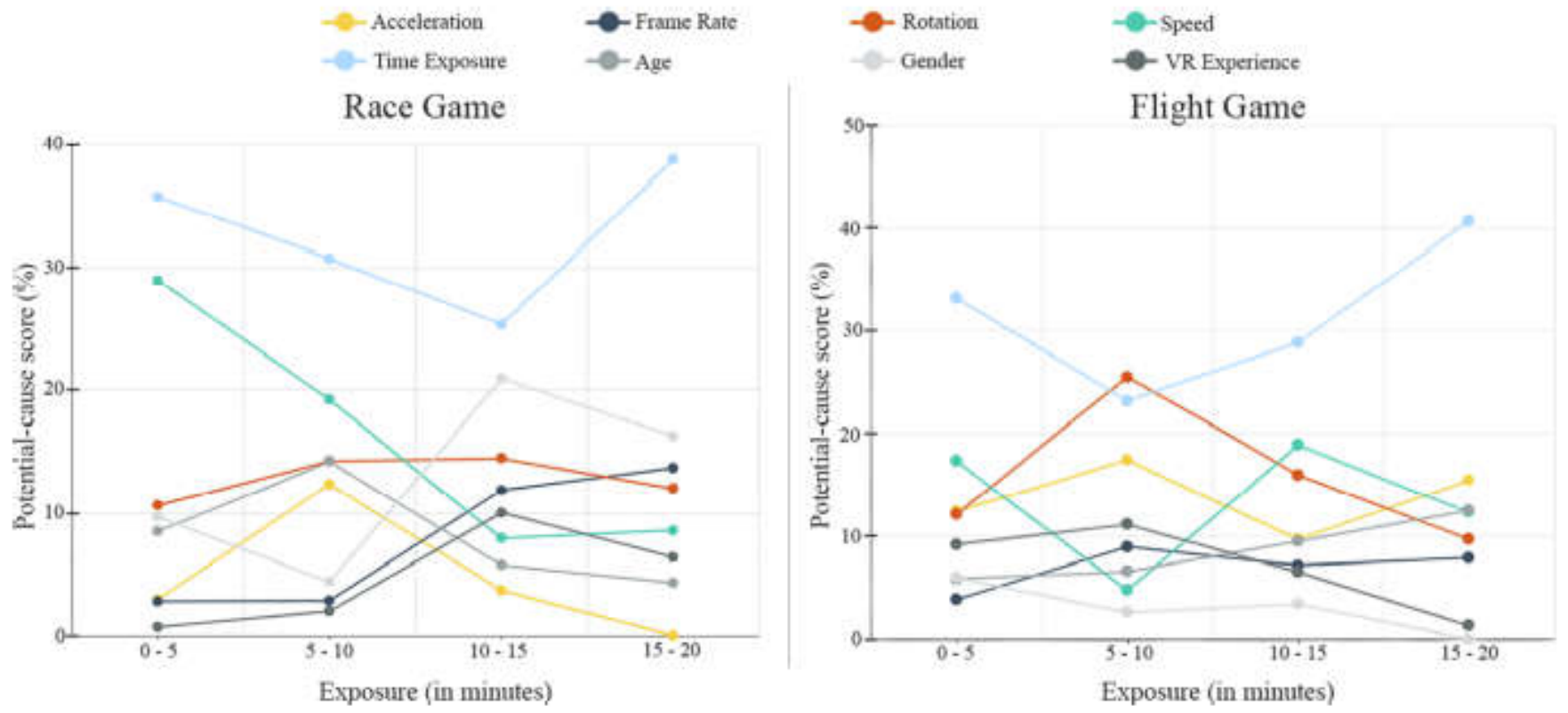

**Fig. 12.** Race and flight game potential-cause score ranking along with different exposure moments for the P6 subjects.

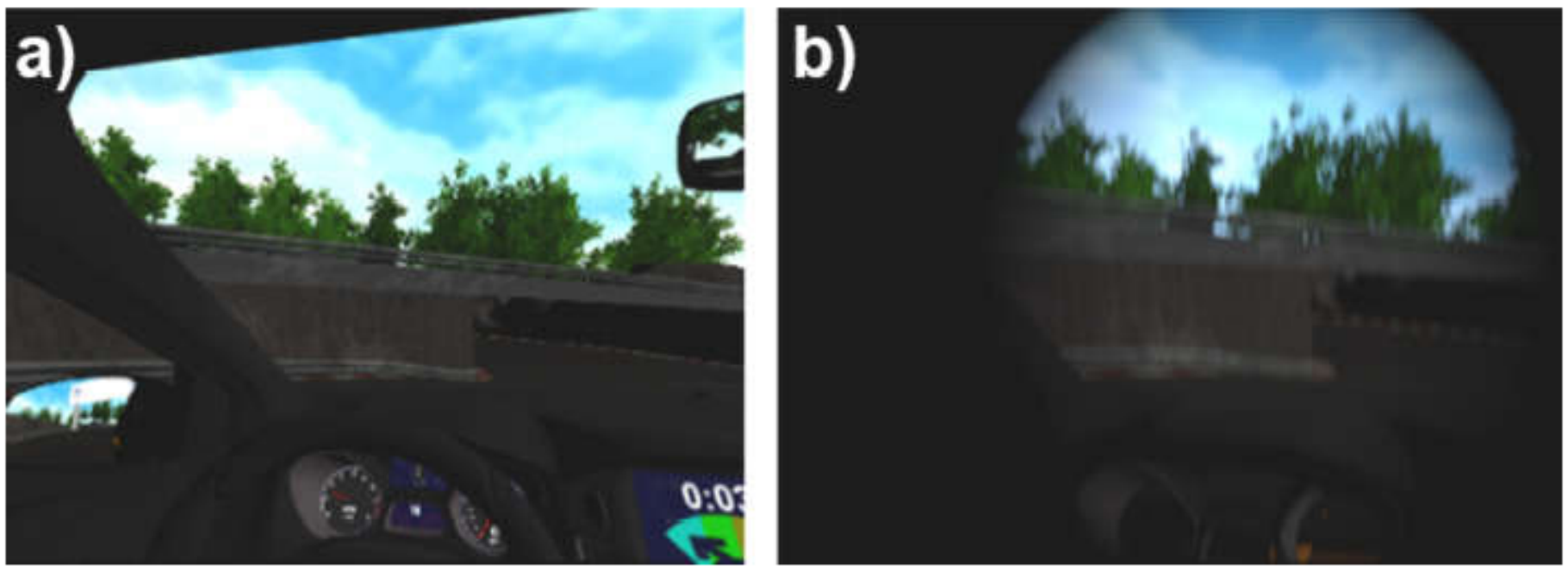

**Fig. 13.** (a) The race gameplay on a curved track with no appliance of strategies. (b) The race gameplay using rotational blurring and tunneling, activated during the gameplay.

### *7.1. Limitations and discussion*

COVID-19 pandemic affected our experiments and protocols in terms of dataset construction. For this reason, some features were also not well represented, such as gender (for women), age (for older adults), and experience (for people with former VR experience). Moreover, the number of used games did not cover locomotion movements, which is specific for games where the user can walk virtually.

Another concern is related to a lack of a more robust treatment of the CS level defined by the verbal feedback of the participants. In other words, verbal feedback is highly subjective and degrees vary from each participant (e.g., moderate for a user may not be moderate for another). Consequently, it is tough to define a robust CS feedback analysis. For this reason, we also considered a binary feedback: no discomfort (for 0 levels of discomfort) and discomfort (for levels 1, 2, and 3).

### *7.2. Future work*

Future work involves including features such as posture, vision impairments, locomotion and other information into our framework. We also aim to improve the balance of the dataset as some cases were also not broadly represented, such as gender (few women), age (few elders) and experience (few subjects with former VR experience). Moreover, as Exposure time was observed as the most frequent cause of CS in our experiments, we also intend to conduct investigative research to detect in which specific timestamp we can identify the increased manifestation of CS symptoms.

Another straightforward way is to explore the gender differences tied to games and virtual reality tasks. Our results and other works [82–84] pointed out that specific tasks can produce different results of discomfort for different user profiles and groups, regarding and not limited to: gender, age, or health issues. Our symbolic machine learning approach can also assist with future analyses.

Moreover, it is necessary to better understand the correlation of profile features with gameplay features, and also how results obtained from profile features (gender, age, VR experience) can be used to label VR experiences according to different groups of users.

Another challenge is to create a virtual reality experience that explores specific tasks individually tied to specific CS causes with a long exposure. The evaluation of individual tasks associated with CS causes may produce a more profound study isolating any other VR potential influences on the CS results.

At last, we would like to highlight that this work can be seen as a preliminary guide to elaborate more adequate game designs for VR games and related applications.

## Declaration of Competing Interest

The authors declare that they have no known competing financial interests or personal relationships that could have appeared to influence the work reported in this paper.